\documentclass[aps,pra,twocolumn,superscriptaddress,showpacs,floatfix,longbibliography,citeautoscript]{revtex4-1}

\usepackage[dvips]{graphicx}

\usepackage{color}

\usepackage{latexsym}

\usepackage{amsmath}

\usepackage{amsthm}

\usepackage{amsfonts}

\usepackage{amssymb}

\usepackage{mathptmx}

\usepackage{subfigure}

\usepackage{mathrsfs}

\usepackage{hyperref}

\hypersetup{colorlinks=true}

\usepackage[all]{hypcap}

\usepackage{gensymb}

\begin{document}
\title{Unusual magnetic structure of high-pressure synthesized perovskites ACrO$_3$(A=Sc, In, Tl)}
\author{Lei Ding}
\email[]{lei.ding@stfc.ac.uk}
\affiliation{ISIS Facility, Rutherford Appleton Laboratory, Harwell Oxford, Didcot OX11 0QX, United Kingdom}
\author{Pascal Manuel}
\affiliation{ISIS Facility, Rutherford Appleton Laboratory, Harwell Oxford, Didcot OX11 0QX, United Kingdom}
\author{Dmitry D. Khalyavin}
\affiliation{ISIS Facility, Rutherford Appleton Laboratory, Harwell Oxford, Didcot OX11 0QX, United Kingdom}
\author{Fabio Orlandi}
\affiliation{ISIS Facility, Rutherford Appleton Laboratory, Harwell Oxford, Didcot OX11 0QX, United Kingdom}
\author{Yu Kumagai}
\affiliation{Materials Research Center for Element Strategy, Tokyo Institute of Technology, Yokohama 226-8503, Japan}
\author{Fumiyasu Oba}
\affiliation{Laboratory for Materials and Structures, Institute of Innovative Research, Tokyo Institute of Technology, Yokohama 226-8503, Japan}
\author{Wei Yi}
\affiliation{Institute of Physics and Beijing National Laboratory for Condensed Matter Physics, Chinese Academy of Sciences, Beijing 100190, China}
\affiliation{International Center for Materials Nanoarchitectonics (WPI-MANA), National Institute for Materials Science (NIMS), Namiki 1-1, Tsukuba, Ibaraki 305-0044, Japan}
\author{Alexei A. Belik}
\affiliation{International Center for Materials Nanoarchitectonics (WPI-MANA), National Institute for Materials Science (NIMS), Namiki 1-1, Tsukuba, Ibaraki 305-0044, Japan}
\affiliation{Research Center for Functional Materials, National Institute for Materials Science (NIMS), Namiki 1-1, Tsukuba, Ibaraki 305-0044, Japan}
\date{\today}

\begin{abstract}
Magnetic structures of metastable perovskites ScCrO$_3$, InCrO$_3$ and TlCrO$_3$, stabilized under high-pressure and high-temperature conditions, have been studied by neutron powder diffraction. Similar to the other orthochromites LnCrO$_3$ (Ln=lanthanide or Y), these materials crystallize into the orthorhombic structure with $Pnma1'$ symmetry. The spin configuration of the metastable perovskites has been found to be $C$-type, contrasting with the $G$-type structure usually observed in LnCrO$_3$. First-principles calculations demonstrate that the $C$-type structure found in ScCrO$_3$ and InCrO$_3$ is attributed to a ferromagnetic (FM) nearest-neighbor interaction, while in TlCrO$_3$, this type of magnetic ordering is stabilized by a strong next-nearest-neighbor antiferromagnetic (AFM) exchange. The spins in the $C$-type magnetic structure line up along the orthorhombic $b$-axis, yielding the $Pnma$ magnetic symmetry. The dominant mechanism controlling this spin direction has been concluded to be the single ion anisotropy imposed by a uniaxial distortion of CrO$_6$ octahedra.
 
\end{abstract}


\maketitle

\section{INTRODUCTION}

Oxide materials with perovskite structure have drawn enormous amounts of interest for several decades due to their central role in fundamental research and technological applications. The remarkable properties of these materials such as colossal magnetoresistance \cite{tokura2000orbital}, superconductivity \cite{Cava1988} and multiferroic behavior \cite{tokura2014, Brink2008} originate from the significant interplay between spin, charge and orbital degrees of freedom.  A key ingredient of this success is the structural versatility of the perovskite lattice which can incorporate a wide range of elements. The diversity of the perovskite materials and their properties has been regarded as a playground to explore novel and unique phenomena. For example, a wide variety of interesting physics was found in the LnMnO$_{3}$ family of manganites, including charge/orbital ordering and metal-insulator transition \cite{tokura2000orbital} as well as multiferroic properties \cite{Bousquet2016, Kimura2003, kenzelmann2005}. Some of Fe$^{3+}$-based perovskites, such as BiFeO$_{3}$ \cite{Wang2003}, GdFeO$_{3}$ \cite{Tokunaga2009} and DyFeO$_{3}$ \cite{Tokunaga2008}, also exhibit coexistence of magnetism and ferroelectricity. The spin-state transition and electronic/ionic conductivity in LnCoO$_{3}$ have been investigated \cite{Itoh2000} for decades. Metal-insulator transition associated with a charge disproportionation has been found in Ni-containing perovskites \cite{Alonso1999, Alonso1999JACS}.  

Another interesting family of perovskite oxides is orthochromites LnCrO$_{3}$, which crystallize into GdFeO$_{3}$-type structure, with the $Pnma1'$ space group.  The magnetic properties of LnCrO$_{3}$ compounds have been extensively investigated \cite{Bertaut1962, Shamir1981, colin2012}. It has been found that these materials undergo AFM transition with a simultaneous weak FM component. Neutron diffraction studies have revealed that they adopt a $G$-type magnetic structure, where all nearest neighbour spins are antiparallel to each other \cite{Bertaut1962, Shamir1981}. The FM component observed in LnCrO$_{3}$ was attributed to a spin canting, allowed by $Pn'm'a$ ($G_y$ -spin configuration) and $Pn'ma'$ ($G_z$ -spin configuration) magnetic symmetry. It was also noted that with a decrease of the radius of the Ln cation, the magnetic ordering temperature was monotonically reduced. Zhou et al. have discussed the relationship between structural distortions and AFM magnetic transition temperature \cite{Zhou2010}. Their work highlights the important role of $t_{2g}-e_g$ hybridization, in addition to the reduction of the orbital overlap integrals. LnCrO$_{3}$ perovskites also show some other interesting physical properties, such as spin-reorientation transition when Ln=Nd, Sm, Gd and Er \cite{Sardar2010}. Recently, orthochromites have attracted revitalised interest due to the observation of a possible multiferroic behavior \cite{Bousquet2016}. 

High pressure synthesis is a well-known technique to expand stability of perovskite structure, which has very recently allowed new members of the Cr$^{3+}$-based perovskites ScCrO$_{3}$, InCrO$_{3}$ and TlCrO$_{3}$ to be investigated \cite{Belik2012, Yi2015, Belik2014}. Similar to LnCrO$_3$, these materials crystallize into orthorhombic $Pnma1'$ symmetry and undergo AFM transition at T$_{N}$=73 K, 93 K and 89 K, respectively. The magnetic transition, however, is not accompanied by a weak FM behavior, reflecting that their magnetic structures might be different from the $G$-type one. In the present study, we explored the magnetic ordering in ACrO$_3$ using neutron diffraction and studied the exchange interactions by means of density functional theory (DFT). 

\section{Technical section}

The polycrystalline samples of ACrO$_3$ were prepared under high pressure and high temperature conditions as described in Refs. \cite{Belik2012, Yi2015}.  The neutron powder diffraction data were collected at the ISIS pulsed neutron and muon facility of the Rutherford Appleton Laboratory (UK), on the WISH diffractometer located at the second target station \cite{Chapon2011}. The samples ($\sim 0.6$ g) were loaded into a cylindrical 3 mm diameter vanadium cans and measured in the temperature range of 1.5 - 110 K using an Oxford Instrument Cryostat. The data at the lowest and the highest temperatures were collected for 1.5 hours, and typical scans between these extremal temperatures were carried out with an exposition time of 20 mins. The temperature step was 10 K and 5 K, in the case of ScCrO$_3$ and InCrO$_3$ perovskites, respectively. For TlCrO$_3$, the data between 1.5 - 80 K were collected with 5 K/step and then the temperature range of 80-95 K was measured with 2-1 K/step and exposition time of 40 mins, to explore the presence of a possible second magnetic transition at T$_{N2}$ = 87 K \citep{Yi2015}; however, no evidence for the second transition was detected. Rietveld refinements of the crystal and magnetic structures were performed using the Fullprof program \cite{Rod1993} against the data measured in detector banks at average $2\theta$ values of 58$^{\circ }$, 90$^{\circ }$, 122$^{\circ }$, and 154$^{\circ }$, each covering 32$^{\circ }$ of the scattering plane. Group-theoretical calculations were done using ISOTROPY \cite{ISOTROPY}, ISODISTORT \cite{ISODISTORT} and Bilbao Crystallographic Server (AMPLIMODES: Symmetry mode analysis \cite{AMPLIMODES} and Magnetic Symmetry and Applications \cite{MAGNDATA}) software.

A series of spin-polarized first-principles calculations was performed for ScCrO$_3$, InCrO$_3$, LuCrO$_3$ and TlCrO$_3$ using the projector augmented-wave (PAW) method \cite{PAW} as implemented in VASP \cite{VASP}.  PAW datasets with radial cutoffs of 1.59, 1.64, 1.48, 1.32, 1.32, and 0.80 $\textup{\AA}$ for Sc, In, Lu, Tl, Cr, and O, respectively, were employed. Sc-$3d$ and -$4s$, In-$5s$ and -$5p$, Lu-$5s$, -$6s$, -$5p$, -$5d$, and -$4f$, Tl-$6s$, -$6p$, and -$5d$, Cr-$4s$ and -$3d$ and O-$2s$ and -$2p$ were described as valence electrons. We adopted the Perdew-Burke-Ernzerhof generalized gradient approximation (GGA) \cite{GGA} to the density functional theory. For correcting on-site Coulomb interactions in the Cr-$3d$ orbitals, we used the $+U$ scheme proposed by Dudarev et al. \cite{Dudarev1998}, in which only $U_{eff} = U - J$ value has a meaning. Since early transition metals have moderate on-site Coulomb potential, we selected $U_{eff}=0$ and 2 eV, the former one is equal to the GGA without the $U$ correction. In all the calculations, lattice constants and internal atomic positions were fully optimized until the residual stresses and forces converged to less than 0.03 GPa and 0.005 eV/$\textup{\AA}$, respectively. Wave functions were expanded with a plane-wave basis set, and the cutoff energies were set to 550 eV. 

\section{Results and discussion}

\begin{figure}[t]
\includegraphics[width=1\linewidth]{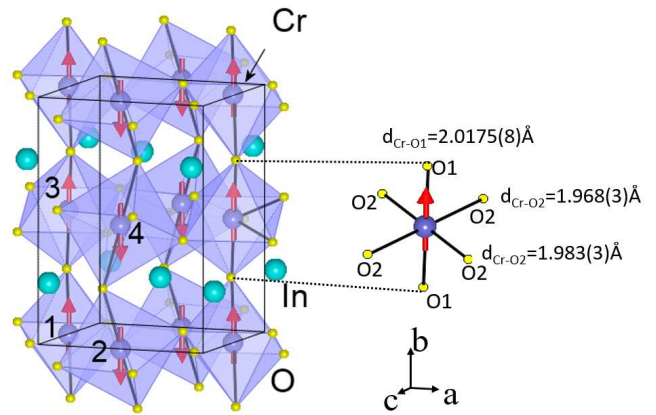}
\caption{(color online) Schematic representation of the crystal and magnetic structures of InCrO$_3$. The four Cr atoms within the unit cell are labeled with 1, 2, 3 and 4. These atoms are used in the definition of the magnetic modes (see Table \ref{tab:1}) and exchange parameters. Cr-O distances were evaluated from the neutron diffraction data collected at 1.5 K.} \label{fig:1}
\end{figure}

\begin{table}[b]
\caption{Structural parameters and Cr-magnetic moments for ACrO$_3$ perovskites refined from the neutron diffraction data collected at 1.5 K in the $Pnma1'$ space group (Wyckoff positions: A-$4c (x,0.25,z)$, Cr-$4b (0,0,0.5)$, O1-$4c (x,0.25,z)$ and O2-$8d (x,y,z)$). Occupancies for all the atoms in the refinement procedure were fixed to the nominal chemical content, since attempts to refine them did not result in any significant deviation from the nominal values and did not improve the fitting quality.}
\centering 
\begin{tabular*}{0.48\textwidth}{@{\extracolsep{\fill}} c c c c c } 
\hline\hline\\  [-2.0ex] 
& Parameter & ScCrO$_3$ & InCrO$_3$ & TlCrO$_3$ \\ 
\hline\\[-2.0ex] 
&$a (\textup{\AA})$ & 5.3613(1) & 5.3539(2) & 5.39539(7)\\ 
&$b (\textup{\AA})$ & 7.3847(2) & 7.5484(3) & 7.6402(2)\\ 
&$c (\textup{\AA})$ & 5.0296(1) & 5.1603(1) & 5.29163(7)\\ 
\hline\\[-2.0ex]
A & $x$ & 0.0709(3) & 0.0615(4) & 0.0466(3)\\ 
& $z$ & 0.9789(3) & 0.9840(5) & 0.9854(2)\\ 
& $B_{iso}(\textup{\AA}^2)$ & 1.72(9) & 1.2(1) & 1.8(7)\\ 
\hline\\[-2.0ex]
Cr & $M(\mu_B)$ & 2.70(2) & 2.50(2) & 2.46(2)\\ 
& $B_{iso}(\textup{\AA}^2)$ & 0.31(7) & 1.2(1) & 1.3(1)\\ 
\hline\\[-2.0ex]
O1 & $x$ & 0.4418(7) & 0.4410(3) & 0.4506(5)\\ 
& $z$ & 0.1294(5) & 0.1239(4) & 0.1094(4)\\ 
& $B_{iso}(\textup{\AA}^2)$ & 1.03(9) & 1.8(1) & 1.4(9)\\ 
\hline\\[-2.0ex]
O2 & $x$ & 0.3072(4) & 0.3083(3) & 0.2999(3)\\ 
& $y$ & 0.0694(2) & 0.0667(2) & 0.0577(2)\\ 
& $z$ & 0.6887(3) & 0.6895(3) & 0.6962(2)\\ 
& $B_{iso}(\textup{\AA}^2)$ & 0.32(9) & 1.6(1) & 1.70(8)\\ 
\hline
\hline  
\end{tabular*}
\label{tab:2} 
\end{table}

The room temperature crystal structures of ACrO$_3$ have been previously investigated using X-ray diffraction \cite{Belik2012, Yi2015}. It has been shown that in all cases, A = Sc, In and Tl, the structure combines both in-phase and out-of-phase octahedral tilting with the $a^+b^-b^-$ tilting pattern in the Glazer's notation \cite{Glazer1972}. This tilted pattern possesses the orthorhombic $Pnma1'$ symmetry and is common for LnCrO$_3$ and many other perovskite systems \cite{Howard1998, Woodward1997}. Our neutron diffraction data revealed that the $Pnma1'$ symmetry of the nuclear structure of ACrO$_3$ does not change down to 1.5 K and a polyhedral representation of the structure is shown in Fig.~\ref{fig:1}. The structural parameters refined at 1.5 K are summarized in Table ~\ref{tab:2}. An important observation coming from the analysis of the structural parameters is the presence of uniaxial distortion of the octahedra, which is most pronounced in the case of InCrO$_3$. The distortion implies that two Cr-O bond distances are notably longer than the other four (Fig.~\ref{fig:1}). This is in contrast with LaCrO$_3$ and other orthochromites with relatively large Ln, in which the octahedra are nearly regular \cite{Zhou2010}. The electronic configuration of the Cr$^{3+}$ ions is not degenerate and the uniaxial elongation of the octahedra in the ACrO$_3$ perovskites studied here is thus a secondary structural effect due to the large octahedral tilting rather than due to Jahn-Teller instability. This structural aspect seems to play a crucial role in the magnetic anisotropy of ACrO$_3$ as discussed below. 

\begin{figure}
\centering
\includegraphics[width=1\linewidth]{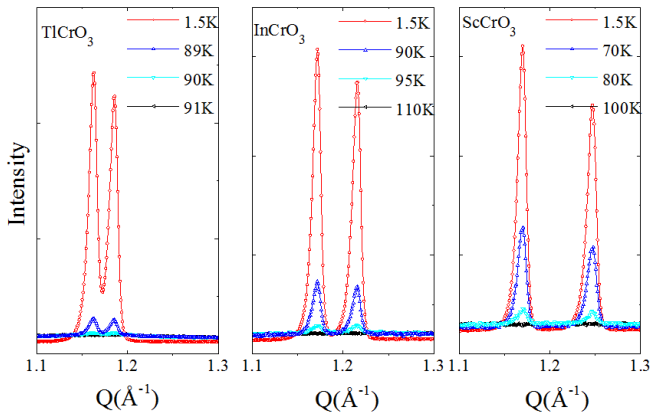}
\caption{(color online) Neutron powder diffraction patterns for ACrO$_{3}$ at selected temperatures, showing the evolution of the two most prominent  magnetic reflections.}\label{fig:2}
\end{figure}

\begin{figure}
\centering
\includegraphics[width=1\linewidth]{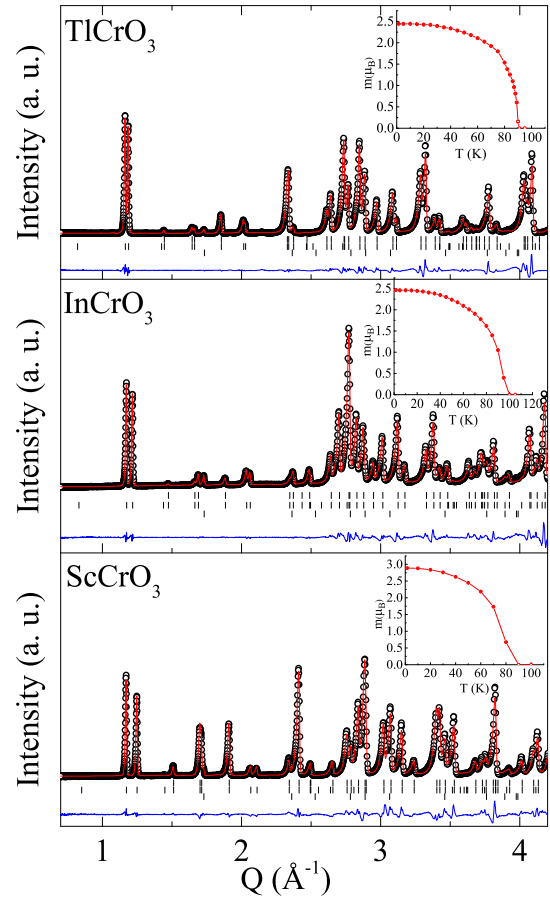}
\caption{(color online) Observed (open circles) and calculated (line) powder neutron diffraction patterns for ACrO$_{3}$ collected at 1.5 K. The nuclear reflections are denoted by upper tick marks. The reflections marked in middle belong to magnetic phase. The lowest tick marks show a Cr$_{2}$O$_{3}$ impurity phase with weight fractions of 3.1(5)\% in ScCrO$_{3}$, 2.1(1)\% in InCrO$_{3}$ and 0.96(3)\% in TlCrO$_{3}$ respectively. The bottom line shows the difference between the observed and calculated diffraction patterns. Inset shows the evolution of the refined Cr$^{3+}$ magnetic moment with temperature.}\label{fig:3}
\end{figure}

Our neutron diffraction data are consistent with the previous magnetic susceptibility and heat capacity measurements, indicating the onset of long-range magnetic ordering below the relevant critical temperatures T$_N$ (Fig.~\ref{fig:2}). In all three cases, Sc-, In- and Tl-based compositions, the magnetic scattering looks very similar (Fig.~\ref{fig:2}), suggesting the same type of the spin ordering. In previous reports \cite{Belik2012, Yi2015}, a possible second transition at lower temperature has been suggested from ac susceptibility measurement. However, our neutron diffraction data demonstrate that no variation of the relative intensities of the magnetic reflections occurs across the second transition, failing to reveal the nature of the low temperature anomalous behavior of the ac susceptibility. All the magnetic reflections can be indexed by $\textbf{k=0}$ propagation vector similar to the perovskites of the LnCrO$_3$ family. A quantitative magnetic structure refinement (Fig.~\ref{fig:3}), however, revealed that it is different from the $G$-type spin configuration (usually found in LnCrO$_3$). 

The refinement procedure governed by the symmetry arguments \cite{Bertaut1968} summarized in Table ~\ref{tab:1} yields the $C_y$-type of magnetic structure, with the $Pnma$ magnetic symmetry. This type of magnetic order implies the spins are ferromagnetically coupled along the $b$-axis and aligned antiferromagnetically in the $(ac)$ plane, as shown in Fig.~\ref{fig:1}. The spin structure is transformed by the $mGM1+$ irreducible representation which also transforms the $G_x$ and $A_z$ magnetic modes. This implies that the latter two are bi-linearly coupled as secondary order parameters to the primary $C_y$ magnetic orderings through the antisymmetric Dzyaloshinskii-Moriya exchange or single ion anisotropy effects. The $G_x$ and $A_z$ spin components, however, are too small to be detected in the powder diffraction experiment. The spin components in the primary $C_y$ magnetic modes were refined to be 2.70(2) $\mu_B$, 2.50(2) $\mu_B$ and 2.46(2) $\mu_B$ at 1.5 K in the case of ScCrO$_3$, InCrO$_3$ and TlCrO$_3$ perovskites, respectively and their temperature dependences are shown in Fig.~\ref{fig:3}. The ordered moments in the latter two are notably smaller than the value expected for $S=3/2$ of Cr$^{3+}$, which might be due to a significant hybridisation of the $d$-orbitals with oxygen or be a result of magnetic fluctuations. The values, however, are in fact very close to the ordered moments found in other orthochromites with non-magnetic $A$-site cations such as LaCrO$_3$ \cite{Martinelli2011}, LuCrO$_3$ \cite{Shamir1981} and Bi$_{1-x}$Y$_x$CrO$_3$ solid solutions \cite{colin2012}. The full results of the magnetic structure refinement can be found in mcif files provided in the Supplemental Material \cite{SM}.

\begin{table}
\caption{Results of the symmetry analysis for the $\bf {k}=0$ magnetic propagation vector and Cr$^{3+}$ cations at the $4b$ position of the parent $Pnma1'$ space group. The four Cr-sites are: Cr1=(0, 0, 0.5), Cr2=(0.5, 0, 1), Cr3=(0, 0.5, 0.5), Cr4=(0.5, 0.5, 1)). Irreducible representations (IRs) are given using the notation of Miller and Love \cite{Miller1967}). Definition of the magnetic modes, using the four Cr-sites, are: F(+~+~+~+),G(+~-~-~+),C(+~-~+~-) and A(+~+~-~-)  \citep{Bertaut1968}. \label{tab:1}}
\begin{tabular}{ccccc}
\hline
\hline
  \multicolumn{1}{c}{IR}& &\multicolumn{1}{c}{Cr(4b)} & & \multicolumn{1}{c}{Space group}  \\
 \hline
 $mGM1+$  & &   ($G_x$,$C_y$,$A_z$)&  & $Pnma$        \\  
 $mGM2+$  & &  ($C_x$,$G_y$,$F_z$)  &  &  $Pn'm'a$            \\
 $mGM3+$ &  & ($F_x$,$A_y$,$C_z$)  & &   $Pnm'a'$           \\
 $mGM4+$ &  & ($A_x$,$F_y$,$G_z$) &  &   $Pn'ma'$ \\
\hline
\hline
\end{tabular}
\end{table}

The magnetic point group, $mmm$, deduced from the present neutron diffraction data is non-polar and does not allow any spontaneous magnetic moment in agreement with the previous magnetization measurements \cite{Belik2012, Yi2015}. In addition, a linear magnetoelectric effect is forbidden as well, casting doubt upon the magnetoelectric nature of the dielectric anomaly at T$_N$, suggested in Refs. \cite{Belik2012, Yi2015}.

\begin{figure}
\includegraphics[width=1\linewidth]{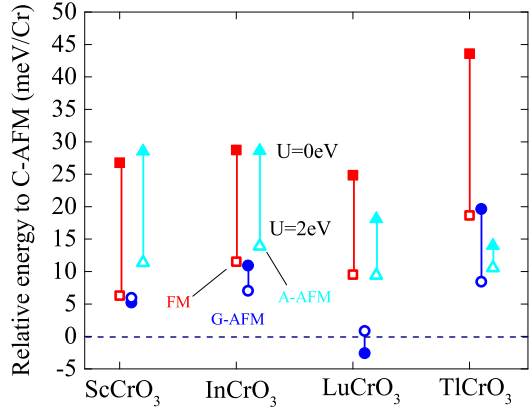}
\caption{(color online) The relative energies of FM, $G$-, and $A$-type AFMs with respect to the energy of $C$-type AFM ordering. The compounds are plotted in order of increasing A-site cation ionic radii, R$_{Sc}$=0.870 $\textup{\AA}$, R$_{In}$=0.92 $\textup{\AA}$, R$_{Lu}$=0.977 $\textup{\AA}$, and R$_{Tl}$=1.027 $\textup{\AA}$, where the modified ionic radius of Tl is used as discussed in Refs.  \cite{Shannon1976,Yi2014}.} \label{fig:4}
\end{figure}

\begin{figure}
\includegraphics[width=1\linewidth]{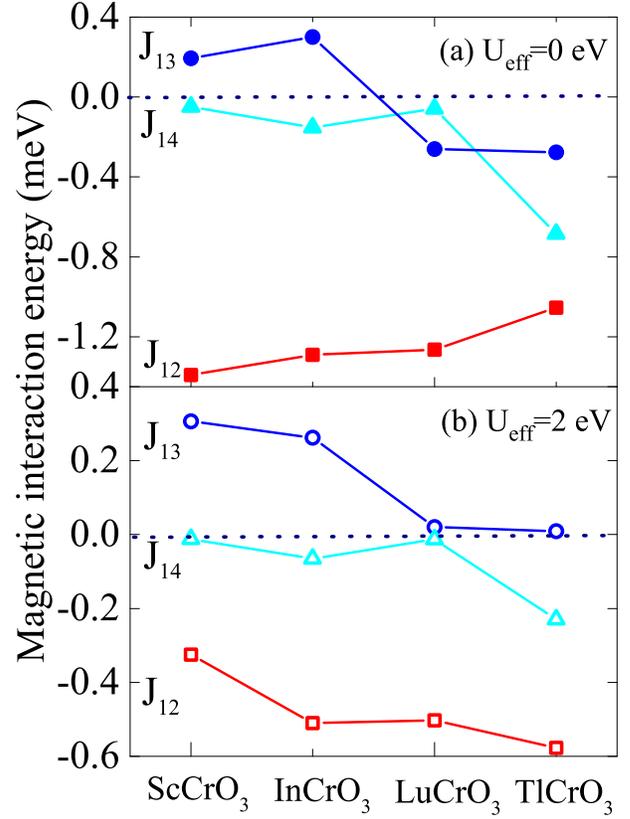}
\caption{(color online) Magnetic interaction energies $J_{12}$, $J_{13}$, and $J_{14}$ calculated using the GGA+$U$ method with $U_{eff}$ = 0 and 2 eV. Note that positive values indicate ferromagnetic interactions, and negative values – antiferromagnetic interactions.} \label{fig:5}
\end{figure}

In order to reveal the factors which underpin the unusual $C$-type magnetic structure in ACrO$_3$, especially in TlCrO$_3$ with the size of Tl$^{3+}$ being larger than that of Lu$^{3+}$ \cite{Yi2015, Shannon1976}, we performed a series of spin-polarized first-principles calculations. LuCrO$_3$, which has the $G$-type magnetic structure experimentally, was chosen as a reference. Four magnetic configurations, i.e., FM and $A$-, $G$-, and $C$-type AFM, were taken into account. Fig.~\ref{fig:4} shows the total energies of FM, $G$- and $A$-type AFM structures relative to that of $C$-type AFM. One can see that the calculated magnetic ground state of ScCrO$_3$, InCrO$_3$, and TlCrO$_3$ is $C$-type AFM structure, in good agreement with the neutron diffraction experiment. On the other hand, the ground state of LuCrO$_3$ is $G$-type AFM structure when $U$ is small, which also agrees with the documented experimental data \cite{Shamir1981}, but the energy becomes closer to that of $C$-type AFM configuration as increases $U$. To understand the origin of the difference in these magnetic ground states, the exchange constants were estimated by mapping the total energy differences between the four magnetic configurations onto the Heisenberg Hamiltonian $H= -2 \Sigma_{i>j}J_{ij} \bf {S_i \cdot S_j}$ \cite{Akamatsu2011,Akamatsu2013}, where $J_{ij}$ is the exchange constant of magnetic interactions between Cr $3d$ spins at the $i$ and $j$ sites (see Fig.~\ref{fig:1}) and $\bf {S_i}, \bf{S_j}$ are the spin vectors at these sites. As the Cr ions are trivalent in ACrO$_3$ and LuCrO$_3$, they have a $d^3$ electronic configuration with $| \bf {S}|$=3/2. In the present analysis, we have considered up to second nearest neighbors for Cr pair interactions. The calculated magnetic interaction energies using $U_{eff}=0$ and 2 eV are shown in Fig.~\ref{fig:5}. $J_{12}$ shows significant AFM interactions in all four compounds. $J_{14}$ is also AFM but it displays remarkable system dependence and is particularly strong in TlCrO$_3$. On the other hand, $J_{13}$ shows notably strong FM interactions in ScCrO$_3$ and InCrO$_3$, whereas it is AFM or very weak FM in LuCrO$_3$ and TlCrO$_3$. These results indicate that the $C$-type AFM structure in ScCrO$_3$ and InCrO$_3$ originates from the FM interactions of $J_{13}$. On the other hand, $C$-type AFM ordering in TlCrO$_3$ is attributed to the strong AFM interactions of $J_{12}$ and $J_{14}$, which results in the FM alignment of Cr1 and Cr3 although $J_{13}$ can have a small AFM interaction. It should be also pointed out that $C$-type magnetic structure has been previously observed in some other perovskite systems such as (Sr/Ca)CrO$_3$ \cite{Komarek2011, Streltsov2008, Lee2009}, LnVO$_3$ \cite{Ren1998, Ulrich2003} and Sr$_{1-x}$Ln$_x$MnO$_3$ \cite{Kurbakov2009, Martin1999}. In all these cases, however, the degenerate electronic configurations of the transition metals lead to spontaneous Jahn-Teller distortions. The $C$-type magnetic structure in these compounds, therefore, can be attributed to anisotropic exchange interactions imposed by the orbital orderings.

\begin{figure}
\includegraphics[width=1\linewidth]{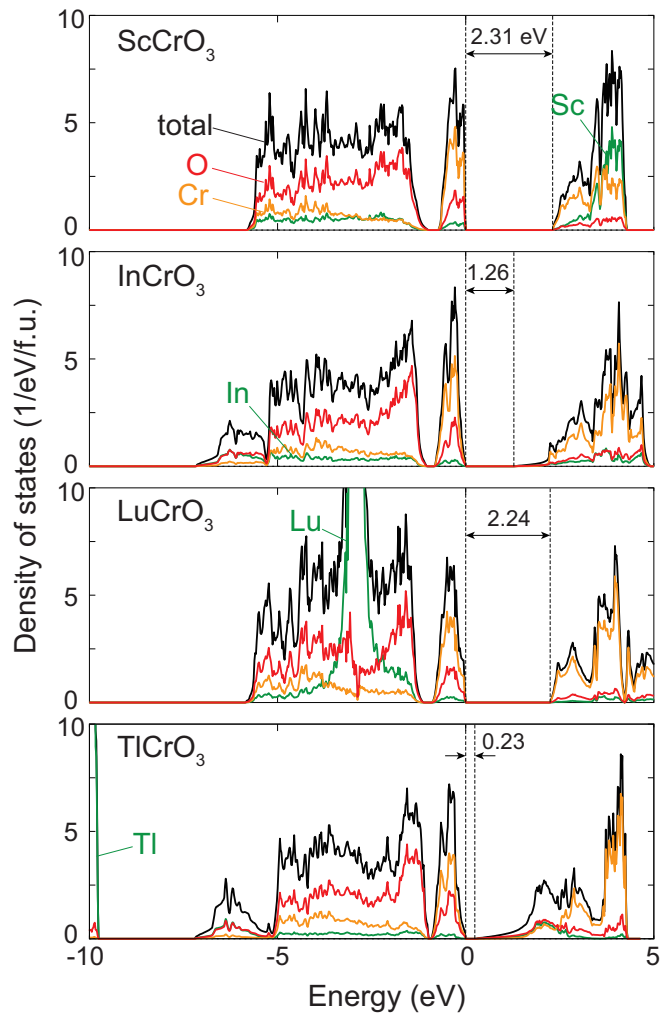}
\caption{(color online) Density of states of the four chromates calculated using GGA+$U$ ($U_{eff}$ = 2 eV). The magnetic configurations are set to the experimentally found ones, namely, $C$-type AFM configuration for ScCrO$_3$, InCrO$_3$, and TlCrO$_3$, and $G$-type one for LuCrO$_3$. Zeros of energy are set to the valence band maxima, and both valence band maxima and conduction band minima are highlighted by dashed lines to guide the eye. The band gaps are also shown.} \label{fig:6}
\end{figure}

Let us briefly discuss the electronic structures of ACrO$_3$ series. Figure~\ref{fig:6} shows the calculated density of states (DOS) of ACrO$_3$ with the experimentally identified magnetic configurations using GGA+$U$ ($U_{eff}$ = 2 eV). Because Cr$^{3+}$ ions are in a $d^3$ configuration, $t_{2g}$ orbitals in the spin-up channel are fully occupied, comprising the valence band maxima in ACrO$_3$. At the same time, the unoccupied $e_g$ orbitals in the same spin channel are located at the bottom of the conduction band, resulting in the (indirect) band gaps in the four orthochromites. Among these, TlCrO$_3$ has a very small gap, which is attributed to the unoccupied Tl-${6s}$ and O-${2p}$ bands. This tendency is the same as those of TlScO$_3$ \cite{Belik2016} and TlMnO$_3$ \cite{Yi2014}. 

The mechanism dominating the magnetic anisotropy of Cr$^{3+}$ in the ACrO$_3$ and related LnCrO$_3$ perovskites is not obvious. Basically, the preferred orientation of the spins in the Cr-sublattice clearly correlates with the size of the $A$-site cation in the perovskite lattice $AB$O$_3$ \cite{Bousquet2016}. In compounds with relatively large Ln=La-Tb, the Cr spins tend to align along the $c$-axis ($a$-axis in the $Pbnm$ setting), like for instance in LaCrO$_3$ and YCrO$_3$, with non-magnetic $A$-site cations. In some cases, with a magnetic rare earth $A$-site sublattice, the direction of the Cr spins changes to be along the $b$-axis (along the $c$-axis in the $Pbnm$ setting) at lower temperatures. These spin reorientation transitions are well-known to be driven by anisotropy of the rare earth cations and the $f-d$ exchange between the $A$- and $B$-site sublattices \cite{Belov1979}. In the perovskites with a smaller Ln=Dy-Lu, the Cr$^{3+}$ spins prefer to order along the $b$-axis. The ScCrO$_3$, InCrO$_3$ and TlCrO$_3$ perovskites follow this empirical correlation, exhibiting the spin configuration with the moments being along the $b$-axis. This change of the magnetic anisotropy seems to correlate with the development of the uniaxial distortion of the CrO$_6$ octahedra in the perovskites with the small $A$-site cations. In LaCrO$_3$, the octahedra are practically isotropic, suppressing the single ion anisotropy. The direction of the spins in this perovskite is believed to be governed by anisotropic antisymmetric exchange \cite{Belov1979}. The $c$-axis is indeed the spin direction which activates the antisymmetric exchange imposed by the largest structural distortions, namely anti-phase and in-phase octahedral tiltings \cite{Khalyavin2014, Khalyavin2015}. However, in the perovskites with smaller $A$-site cations and distorted octahedra, the single ion anisotropy prevails, forcing the spins to be along the $b$-axis. The uniaxial elongation of octahedron lifts the degeneracy between the half-occupied $xy$, $xz$ and $yz $ orbitals of Cr$^{3+}$, increasing the energy of the former and decreasing the energy of the latter two. In this case, the single ion anisotropy is assumed to be dominated by the spin-orbit interaction between the occupied $xy \uparrow $ and empty $x^2-y^2 \uparrow $ electronic states, resulting in the easy axis type of anisotropy as suggested in the recent perturbation theory by Whangbo et al. \cite{Whangbo2015}. This is the case of a non-degenerate perturbation with a weak magnetic anisotropy. Thus, locally, the easy axis of the Cr$^{3+}$ spin is along the direction of the uniaxial elongation of octahedron. The local easy axes are tilted away from the $b$-direction, following the octahedral rotations (Fig.~\ref{fig:1}). When this type of anisotropy is combined with the strong symmetric superexchange interactions, requiring  a collinear spin alignment, it yields the $b$-axis as the global easy axis of the nearly collinear magnetic structure. Finally, let us note that the swapping from the $G_z$ spin configuration in YCrO$_3$ to the $G_y$ one in LuCrO$_3$ due to a single ion anisotropy effect has been also proposed in the early work by Kadomtseva et al. \cite{Kadomtseva1981} based on the study of spin reorientation induced by external magnetic fields.

\section{Conclusion}

The high-pressure stabilized metastable perovskites ScCrO$_3$, InCrO$_3$ and TlCrO$_3$ exhibit a long-range AFM order below T$_N \sim$ 73 K, 93 K and 89 K, respectively. The primary spin configuration in all cases is $C_y$-type, where the spins are ferromagnetically coupled along the $b$-axis and antiferromagnetically along both $a$- and $c$-directions. In ScCrO$_3$ and InCrO$_3$, this magnetic structure is stabilized by the FM nearest-neighbour exchange interaction along the $b$-axis and AFM in the $(ac)$ plane, while in TlCrO$_3$, it is due to a strong AFM next-nearest-neighbour exchange. The direction of the spins in the primary magnetic mode is governed by a single ion anisotropy imposed by uniaxial elongation of the CrO$_6$ octahedra. It seems that antisymmetric exchange and single ion effects are in general the dominant mechanisms of the magnetic anisotropy of the Cr-sublattice in orthochromite perovskites with relatively large and small $A$-site cations, respectively. 
 
\section{Acknowledgments}

L. D. thanks support from the Rutherford International Fellowship Programme (RIFP). This project has received funding from the European Union's Horizon 2020 research and innovation programme under the Marie Sk\l{}odowska-Curie grant agreements No.665593 awarded to the Science and Technology Facilities Council. P. M. and D. D. K. acknowledge a support from the project TUMOCS. This project has received funding from the European Union's Horizon 2020 research and innovation programme under the Marie Sk\l{}odowska-Curie grant agreements No.645660. The work done at NIMS was supported by World Premier International Research Center Initiative (WPI Initiative, MEXT, Japan). The work done at Tokyo Institute of Technology was supported by the Grants-in-Aid for Young Scientists A (Grant No.15H05541).

\end{document}